\documentclass[3p,twocolumn]{elsarticle}

\journal{Journal of \LaTeX\ Templates}

\usepackage{comment}
\usepackage{graphicx}
\usepackage[colorinlistoftodos]{todonotes}
\usepackage{multirow}
\usepackage{mathrsfs}
\usepackage{url}
\usepackage{empheq}  
\usepackage{amssymb}
\usepackage{txfonts}
\usepackage{mathtools}
\usepackage{siunitx}
\usepackage{tabularx}
\usepackage{array}
\usepackage{microtype}
\usepackage[colorlinks=true, allcolors=blue]{hyperref}

\DeclarePairedDelimiter{\abs}{\lvert}{\rvert}

\begin{document}

\begin{frontmatter}

\title{Magnetic field screening in strong crossed electromagnetic fields}

\author[icra,icranet]{S.~Campion}

\author[icra,icranet,icranetferrara,unife,inaf1]{J.~A.~Rueda\corref{mycorrespondingauthor}}
\cortext[mycorrespondingauthor]{Corresponding author}
\ead{jorge.rueda@icra.it}

\author[icra,icranet,nizza,inaf2]{R.~Ruffini}

\author[icra,icranet]{S.-S.~Xue}

\address[icra]{ICRA, Dipartimento di Fisica, Sapienza Universit\`a  di Roma, P.le Aldo Moro 5, 00185 Rome, Italy}
\address[icranet]{International Center for Relativistic Astrophysics Network, Piazza della Repubblica 10, I-65122 Pescara, Italy}
\address[icranetferrara]{ICRANet-Ferrara, Dipartimento di Fisica e Scienze della Terra, Universit\`a degli Studi di Ferrara, Via Saragat 1, I--44122 Ferrara, Italy}
\address[unife]{Dipartimento di Fisica e Scienze della Terra, Universit\`a degli Studi di Ferrara, Via Saragat 1, I--44122 Ferrara, Italy}
\address[inaf1]{INAF, Istituto de Astrofisica e Planetologia Spaziali, Via Fosso del Cavaliere 100, 00133 Rome, Italy}
\address[nizza]{Universit\'e de Nice Sophia Antipolis, CEDEX 2, Grand Ch\^{a}teau Parc Valrose, Nice, France}
\address[inaf2]{INAF, Viale del Parco Mellini 84, 00136 Rome, Italy}

\begin{abstract}
We consider crossed electric and a magnetic fields $\left(\vec{B}=B\,\hat{z},~\vec{E}=E\,\hat{y}\right)$, with $E/B<1$, in presence of some initial number of $e^{\pm}$ pairs. We do not discuss here the mechanism of generation of these initial pairs. The electric field accelerates the pairs to high-energies thereby radiating high-energy synchrotron photons. These photons interact with the magnetic field via magnetic pair production process (MPP), i.e. $\gamma+B\rightarrow e^{+}+e^{-}$, producing additional pairs. We here show that the motion of all the pairs around the magnetic field lines generates a current that induces a magnetic field that shields the initial one. For instance, for an initial number of pairs $N_{\pm,0}=10^{10}$, an initial magnetic field of $10^{12}$~G can be reduced of a few percent. The screening occurs in the short timescales $10^{-21}\leq t \leq 10^{-15}$~s, i.e. before the particle acceleration timescale equals the synchrotron cooling timescale. The present simplified model indicates the physical conditions leading to the screening of strong magnetic fields. To assess the occurrence of this phenomenon in specific astrophysical sources, e.g. pulsars or gamma-ray bursts, the model can be extended to evaluate different geometries of the electric and magnetic fields, quantum effects in overcritical fields, and specific mechanisms for the production, distribution, and multiplicity of the $e^{-}e^{+}$ pairs.
\end{abstract}


\end{frontmatter}


\section{Introduction}
\label{sec1}
%
The process of screening of a strong electric field by means of the creation of electron-positron ($e^{\pm}$) pairs through quantum electrodynamics (QED) particles showers has been studied for many years. Recently, it was shown in~\cite{fedotov2010limitations} that an electric field as high as $E\sim\alpha_f E_{cr}$, where $\alpha_f$ is the fine structure constant and $E_{cr}=m_e^2 c^3/(e\hbar)\approx 1.32\times 10^{16}$~V/cm is the critical field for vacuum polarization (see \cite{2010PhR...487....1R} for a review), cannot be maintained because the creation of particle showers depletes the field. To the best of our knowledge, no analog conclusion has been reached for a magnetic field. The main topic of this paper is to build a simple model to analyze the magnetic field screening (MFS) process owing to the motion of $e^{\pm}$ pairs in a region filled by magnetic $\vec{B}$ and electric $\vec{E}$ fields. 

The basic idea of the screening process is explained by the following series steps:
\begin{enumerate}
\item 
An initial number of $e^{\pm}$ is placed in a region filled by $\vec{E}$ and $\vec{B}$, with $E/B <1$, $B\leq B_{cr} = m_e^2 c^3/(e\hbar) \approx 4.4\times 10^{13}$~G and then $E<E_{cr}$. These initial pairs could have been the result of vacuum breakdown, but 
we do not discuss here their creation process.
\item 
The initial pairs are accelerated by $\vec{E}$ and emit radiation via the curvature/synchrotron mechanism (or their combination), due to the $\vec{B}$ field.
\item 
The photons create a new $e^{\pm}$ pairs via the magnetic pair production process (MPP), $\gamma+B\rightarrow e^{-} + e^{+}$.
\item 
Also these new pairs are accelerated, radiate photons and circularize around the magnetic field lines. This circular motion generates a current that induces a magnetic field, $\vec{B}_{ind}$, oriented in the opposite direction with respect to the original one, thereby screening it. Due to the creation of new charged particles and to the proportionality between the strength of the fields, also the electric field is screened.
\item 
Since the series of the previous processes occurs at every time $t$, they could develop a particle shower.
\end{enumerate} 

A possible astrophysical scenario in which this study finds direct application is in the process of high-energy (MeV and GeV) emission from a BH in long gamma-ray bursts (GRBs), in view of the recently introduced ``\textit{inner engine}'' \cite{2019ApJ...886...82R, 2020EPJC...80..300R} in the binary-driven hypernova (BdHN) model (see, e.g., \cite{2019ApJ...871...14B, 2019ApJ...874...39W, 2020ApJ...893..148R}). The \textit{inner engine} is composed of the newborn rotating BH, surrounded by the magnetic field (inherited from the collapsed NS) and low-density ionized matter from the SN ejecta, and it is responsible of the high-energy (GeV) emission observed in the GRB. The gravitomagnetic interaction of the rotating BH and the magnetic field induces an electric field which accelerates $e^-$ which emit GeV photons by synchrotron radiation \cite{2019ApJ...886...82R}. 
It has been argued in \cite{2020ApJ...893..148R} that the magnetic field surrounding the BH could exceed the critical value, i.e. $B>B_{cr}$. Therefore, a situation in which $E\gtrsim E_{cr}$ could occur leading to a vacuum polarization process \cite{2010PhR...487....1R}. This could be the seed of the $e^{\pm}$ pairs we start with. 
If the MFS
occurs, the optical depth for synchrotron photons could decrease sufficiently to allow them to freely escape from the region near the BH and become observable. Therefore, the physical process that we present here could be necessary to lead to the astrophysical conditions derived in \cite{2019ApJ...886...82R} for the explanation of the GeV emission observed in long GRBs

In this article, we build a first, simplified framework to study the problem of the MFS 
by $e^{\pm}$ pairs. 
We analyze the whole screening process for the specific configuration of perpendicular fields: $\vec{E}=E~\hat{y}$ for the electric field and $\vec{B}=B~\hat{z}$ for the magnetic field.

%
%
\section{Particles dynamics}
\label{sec2}
%
In this section, we start to build the equations that describe the particles dynamics and their creation.

The equations of motion of a particle immersed in an EM field\footnote{We use throughout cgs-Gaussian units in which the magnetic and electric fields share the same dimensions (g$^{1/2}$~cm$^{-1/2}$~s$^{-1}$). We also use a $-2$ signature so the spacetime metric is $\eta_{\mu\nu}=(1,-1,-1,-1)$.} read (see, e.g.,~\cite{landau1971classical, jackson1999classical}) 
\begin{subequations}\label{12}
\begin{align}
\frac{d\vec{r}}{dt}&=c\vec{\beta},\\
\frac{d\vec{\beta}}{dt}&=\frac{e}{mc\gamma}\left[\vec{E}+\vec{\beta}\times \vec{B} -\vec{\beta}(\vec{E}\cdot\vec{\beta})\right],\\
\frac{d\gamma}{dt}&=\frac{e}{mc}\left(\vec{E}\cdot\vec{\beta}
\right)-\frac{I}{mc^2}, \label{6c}
\end{align}
\end{subequations}
where $I$ is the energy loss per unit time due to the radiation emitted by an accelerated particle. 
Following \cite{kelner2015synchro}, we use the energy loss in the quantum regime written as:
\begin{equation}
\label{15}
I\equiv\left\lvert-\frac{d E}{dt}\right\rvert=\frac{e^2~m^2~c^3}{\sqrt{3}~\pi~\hslash^2} \overline{H}(\chi),
\end{equation}
with $\overline{H}(\chi)$ defined in~\cite{kelner2015synchro}.
The parameter $\chi$ is defined as $\chi\equiv\varepsilon_{*}/2~\varepsilon_e$ (see~\cite{kelner2015synchro}, and references therein for details), where $\varepsilon_{*}=\hslash \omega_{*}$ is the \textit{critical photons energy,} with: 
\begin{equation}
\label{14}
\omega_{*}=\frac{3 e \gamma^2}{2 mc}\sqrt{\left(\vec{E}+\vec{\beta}\times\vec{B}\right)^2-\left(\vec{\beta}\cdot\vec{E}\right)^2},
\end{equation}
and being $\varepsilon_e=\gamma m_e c^2$ the electron/positron energy. For $\chi\gtrsim 1$, the particle radiates in the so-called quantum regime while, for $\chi<1$, the particle radiates in the classical regime (see also Section~\ref{subsecfurtherMPPcond}). Equation~\eqref{15} 
 is valid in both regimes.

The radiation emitted by an accelerating particle with Lorentz factor $\gamma$ is seen by an observer at infinity as confined within a cone of angle $\sim 1/\gamma$. Therefore, for ultra-relativistic particles ($\gamma \gg 1)$, the radiated photons are seen to nearly follow the particle's direction of motion. We denote by $\phi$ the angle between the particle/photon direction and the magnetic field. Consistently, the square root in Eq.~\eqref{14} already takes into account the relative direction between the photons and the fields.

The screening process starts when electrons are emitted inside the region where both $\vec{E}$ and $\vec{B}$ are present and proceeds through the series of steps described in section~\ref{sec1}.
The evolution with time of the photon number can be written as
\begin{equation}
\label{17}
\frac{dN_{\gamma}}{dt}(t,\phi)=N_{\pm}(t,\phi)\,\frac{I(t)}{\varepsilon^{e}_{\gamma}(t)},
\end{equation}
where $I$ is the intensity in Eq.~\eqref{15} and $N_{\pm}$ is the number of created pairs via the MPP process.
The number of pairs is strictly related to the number of photons. Then, the equation for the evolution of the number of created pairs $N_{\pm}$ can be written as
\begin{equation}
\label{18}
\frac{d N_{\pm}}{d t}(t,\phi)=N_{\gamma}\left(t\right) R^e_A\left(t,\phi\right)c,
\end{equation}
%
where $R^{e}_{A}$ is the attenuation coefficient for the MPP process (see section~\ref{sec4}).
%
%
\section{Magnetic field equation}
\label{sec3}
%
Let us introduce the curvature radius of the particle's trajectory \cite{kelner2015synchro}
\begin{equation}
\label{19}
\frac{1}{R_c}=\left\lvert\frac{d\vec{\beta}}{cdt}\right\rvert=\frac{e}{\gamma m c^2}\sqrt{\left(\vec{E}_{tot}+\vec{\beta}\times\vec{B}_{tot}\right)^2-\left(\vec{\beta}\cdot\vec{E}_{tot}\right)^2},
\end{equation}
where $B_{tot}$ and $E_{tot}$ are the total magnetic and electric fields, respectively, as defined below.

The motion of a particle in the present EM field can be considered as the combination between acceleration along the $z-$direction, and in a series of coils around the magnetic field lines, in the $x-y$ plane. The linear number density of the particles on a path $dl$ is defined as $n_{\lambda}=dN_{\pm}/dl$, while the current density in the two directions are $\vec{J}_{\perp}=e~\vec{\beta}_{\perp}\, n_{\lambda}\, c$ and $\vec{J}_{\parallel}=e~\vec{\beta}_{\parallel}\, n_{\lambda}\, c$,
with $\beta_{\perp}=(~\beta_x^2+\beta_y^2~)^{1/2}$ and $\beta_{\parallel}=\beta_z$.

\begin{figure}[ht!]
\centering
\includegraphics[width=0.6\hsize,clip]{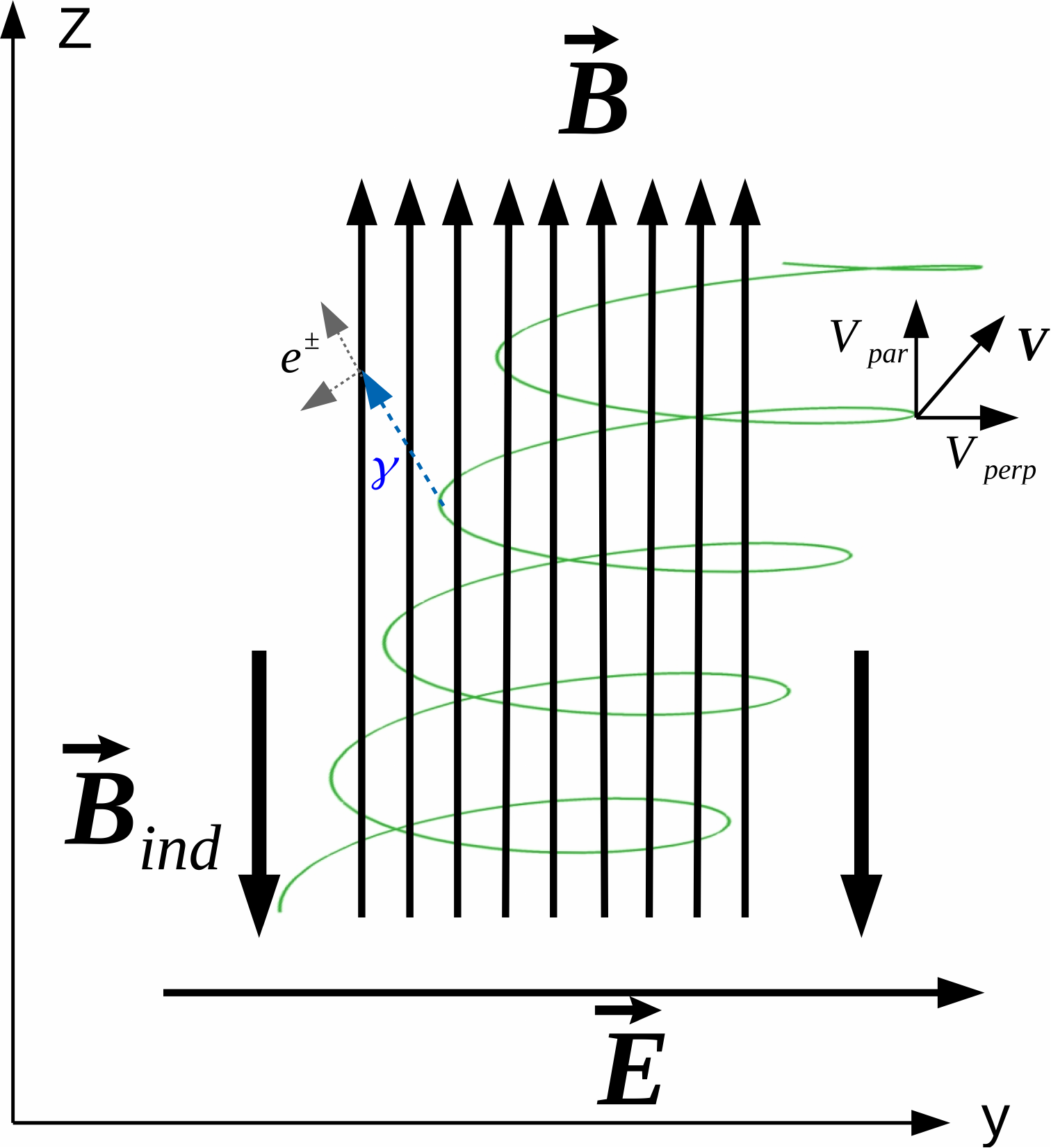}
\caption{Schematic representation of the screening process for perpendicular fields ($\vec{B}=B\,\hat{z},~\vec{E}=E\,\hat{y}$). The green lines represent the coils of the particles path.}
\label{figscreeningscheme}
\end{figure}

The infinitesimal induced magnetic field $d\vec{B}_{ind}$ generated by the current of an element of the coil $dl=\lvert d\vec{l}~\rvert$ is:
\begin{equation}
\label{21}
d\vec{B}_{ind}=\frac{J_{\perp}}{c}\frac{d\vec{l}\times\Delta \vec{r}}{\left\lvert\Delta \vec{r}~\right\rvert^{~3}}=\frac{J_{\perp}}{c}\frac{\left\lvert d\vec{l}~\right\rvert}{\left\lvert\Delta \vec{r}~\right\rvert^{~2}} ~\hat{n}=e~\beta_{\perp}  \frac{dN_{\pm}}{dl}\frac{dl}{\left\lvert\Delta \vec{r}~\right\rvert^{~2}}~\hat{n},
\end{equation}
where $\Delta \vec{r}$ is the vector connecting an element of the coil, in the $x-y$ plane, with an element of the coil axes and $\hat{n}$ is the versor normal to the $\Delta\vec{r}-d\vec{l}$ plane (since $d\vec{l}$ and $\Delta \vec{r}$ are always perpendicular). 
The only non-zero component of the magnetic field vector is the one parallel to the coil axes. Then, we have only $dB_z=d B\sin\theta$, where $\sin(\theta)=R_c(t)/\left\lvert\Delta \vec{r}~\right\rvert$ and $\left\lvert\Delta \vec{r}~\right\rvert=\sqrt{z^2+R_c(t)^2}$, with $z$ the height on the coil axes. 
At the coil center ($z=0$) and writing $dl=c~dt$, we obtain
\begin{equation}
\label{22}
\frac{dB_{z,ind}}{dt}=e\frac{\beta_{\perp}(t)}{R_c(t)^2}\frac{dN_{\pm}}{dt}.
\end{equation}
Here $B_{tot}(t)=B_0-B_{ind}(t)$ is the total magnetic field;
$B_0$ is the initial background magnetic field.
Figure~\ref{figscreeningscheme} shows a schematic representation of the screening process. 
%

%
\section{Pair production rate}
\label{sec4}
%
In Eq.~\eqref{18}, we introduced the attenuation coefficient for the magnetic pair production $R^{e}_A$. 
Hereafter, we refer to as $\zeta\equiv R^e_A c$ the MPP rate. 
In~\cite{daugherty1975pair}, it was derived the expression for the pair production rate, in the observer frame at rest, in strong perpendicular electric and magnetic fields ($\vec{E}\cdot\vec{B}=0$).

%
\subsection{Production rate for perpendicular fields}
\label{subsec4.1}
%
In the present configuration of the fields, we study the pair production for a general direction propagation of photons. Let us consider a photon with energy $\varepsilon_{\gamma}$ and momentum vector $\hslash \vec{k}$, with director cosines $\left(\eta_x,~\eta_y,~\eta_z\right)$.
Following~\cite{daugherty1975pair}, we apply a Lorentz transformation along the $x-$direction to a new frame, $K^{'}$, where there is no electric field; we calculate all the necessary quantities and the rate in $K^{'}$ and, finally, we transform them back to the lab frame.

We introduce the photon four-momentum as $k^{~\mu}=(\omega/c, \vec{k})$ and the four-vector for the photon direction as $\eta^{~\mu}=(1, \vec{k}/k^0)$, with $k^{0}$ the time component of $k^{~\mu}$, i.e. the photon energy. 
The photon energy and director cosines in the $K^{'}$ frame are ( $1$, $2$, $3$ stand for $x$, $y$, $z$):
\begin{subequations}\label{26}
\begin{align}
{\varepsilon^{'}}_{\gamma}&=\gamma^{*} \left(1-\beta^{*} \eta^{1}\right) \varepsilon_{\gamma}\\
{\eta^{'}}^{1}&=\frac{k^0}{k^{'}_0} \Lambda^{1}_{\nu} \eta^{~\nu}=\frac{\varepsilon_{\gamma}}{\varepsilon^{'}_{\gamma}}\gamma^{*}\left(\eta^1-\beta^{*}\right)\\
{\eta^{'}}^{2}&=\frac{\varepsilon_{\gamma}}{\varepsilon^{'}_{\gamma}} \eta^{2},~ {\eta^{'}}^{3}=\frac{\varepsilon_{\gamma}}{\varepsilon^{'}_{\gamma}} \eta^{3},
\end{align}
\end{subequations}
where $\varepsilon_{\gamma}=\hslash k^0$. The component of the magnetic field in the $K^{'}$ frame perpendicular to the propagation direction of the photons, is given by:
\begin{equation}
\label{27}
\vec{B^{'}}\times \vec{\eta^{'}}=\left(B^{'}_{\parallel} \hat{e^{'}}_{\parallel}+B^{'}_{\perp} \hat{e^{'}}_{\perp}\right)\times \hat{e^{'}}_{\parallel}=B^{'}_{\perp} \left(\hat{e^{'}}_{\perp}\times \hat{e^{'}}_{\parallel}\right)=B^{'}_{\perp},
\end{equation}
where $\hat{e^{'}}$ are the basis vectors of the $K^{'}$ frame. 
The vector $\vec{B^{'}}_{\perp}=\left(-B^{'}_z \eta^{'}_y,\,B^{'}_z \eta^{'}_x\right)$ and then, from Eq.~\eqref{26}, we get the magnitude of $B^{'}_{\perp}$ as a function of the fields, the photon director cosines and energy in the laboratory frame:
\begin{equation}
\label{28}
B^{'}_{\perp}=B_z \sqrt{1-\frac{E_y^2}{B_z^2}} \frac{\varepsilon_{\gamma}}{\varepsilon^{'}_{\gamma}} \sqrt{\eta_y^2+{\gamma^{*}}^2 \left(\eta_{x}-\beta^{*}\right)^2}.
\end{equation}
The pair production rate in the $K^{'}$ frame is given by \cite{daugherty1975pair}:
\begin{subequations}
\label{29}
\begin{align}
\zeta^{'} &=0.23 \frac{\alpha_f c}{\lambdaslash_c} \frac{B^{'}_{\perp}}{B_{cr}} \exp\left(-\frac{4}{3}\Psi^{-1}\right), \label{29a}\\
\Psi &= \frac{1}{2}\left(\frac{\varepsilon^{'}_{\gamma}}{mc^2}\right)\left(\frac{B^{'}_{\perp}}{B_{cr}}\right). \label{29b}
\end{align}
\end{subequations}
The expression for the rate in Eq.~\eqref{29a} is valid as long as $\Psi\ll 1$ (see below section~\ref{subsecinitialconditions}).

The pair production rate in the laboratory frame, $K$ (observer at infinity), is given by $\zeta=\zeta^{'}/\gamma^{*}$, that can be rewritten as a function of the variables in the $K$ frame as
\begin{equation}
\label{31}
\begin{split}
\zeta=0.23 \frac{\alpha_f c}{\lambdaslash_c} \frac{B_z}{B_{cr}} \left(1-\frac{E_y^2}{B_z^2}\right)\frac{\sqrt{\eta_y^2\left(1-\frac{E_y^2}{B_z^2}\right)+\left(\eta_x-\frac{E_y}{B_z}\right)^2}}{1-\frac{E_y}{B_z}\eta_x}\\
\times\exp\left\{-\frac{8}{3}\frac{mc^2}{\varepsilon_{\gamma}}\frac{B_{cr}}{B_z}\left[\eta_y^2\left(1-\frac{E_y^2}{B_z^2}\right)+\left(\eta_x-\frac{E_y}{B_z}\right)^2\right]^{-1/2}\right\}.
\end{split}
\end{equation}

One can write the photon momentum director cosines $\vec{\eta}$ as a function of the electron velocity $\vec{\beta}$, the polar $\Theta$ and azimuthal $\Phi$ angles of emission in the comoving frame: 
\begin{subequations}\label{33}
\begin{align}
\eta_x &=\frac{\sin\Theta\cos\Phi+\beta_x\left[\gamma+\frac{(\gamma-1)}{\beta^2}\nu\right]}{\gamma\left(1+\nu\right)}\\
\eta_y &=\frac{\sin\Theta\sin\Phi+\beta_y\left[\gamma+\frac{(\gamma-1)}{\beta^2}\nu\right]}{\gamma\left(1+\nu\right)}\\
\eta_z &=\frac{\cos\Theta+\beta_z\left[\gamma+\frac{(\gamma-1)}{\beta^2}\nu\right]}{\gamma\left(1+\nu\right)},
\end{align}
\end{subequations}
where $\nu=\beta_x \sin\Theta\cos\Phi+\beta_y\sin\Theta\sin\Phi+\beta_z\cos\Theta$ and $\gamma$ the $e^{\pm}$ Lorentz factor. 
Selecting specific photons emission angles in the comoving frame (e.g. $\Theta=\Phi=\pi/2$), we can now integrate our set of equations.
%
\section{Results}
\label{sec6}
%
We now present the results of the numerical integration of the set of equations described in the previous sections, with the related initial conditions (hereafter ICs).
In our calculations, we adopt the electric and magnetic field strengths proportional to each other, i.e.:
\begin{equation}
\label{42}
E(t)=\Upsilon~B(t),
\end{equation}
where $0< \Upsilon \leq 1$ since 
we are interested in analyzing situations of magnetic dominance. We have selected 
three values of reference, $\Upsilon=1/2$, $1/10$, and $1/100$. The proportionality is requested at any time, so when $B(t)$ changes, $E(t)$ changes accordingly to keep $\Upsilon$ constant. These combined effects affect the motion of particles and, consequently, all the successive processes giving rise to the screening.

\subsection{Initial conditions and MPP rate}
\label{subsecinitialconditions}
%
In order to apply Eq.~\eqref{31}, 
the condition $\Psi\ll 1$ (expressed in the $K^{'}$ frame) must be satisfied. 
 Transforming back $\epsilon^{'}_{\gamma}$ and $B^{'}_{\perp}$ to the original $K$ frame (where both fields are present), we obtain the following condition for $\Psi$: 
\begin{equation}
\begin{split}
\label{45} 
\Psi=\frac{3}{4}\frac{e \hslash}{m~c}&\frac{B^2}{B_{cr}}\gamma^2 \sqrt{{\beta_y}^2\left(1-\frac{E^2}{B^2}\right)+\left(\frac{E}{B}-\beta_x\right)^2}\\
& \times \sqrt{{\eta_y}^2 \left(1-\frac{E^2}{B^2}\right)+\left(\eta_x -\frac{E}{B}\right)^2}\ll 1.
\end{split}
\end{equation}
This condition brings with it three conditions for the initial values of the variables: 
$B_0,~\gamma_0$, 
and particles direction of emission (contained in the initial velocities $\vec{\beta}_0$ 
and in the director cosines of the photons~$\vec{\eta}$). Then, we need to choose the right values for the three parameters in order to apply Eq.~\eqref{31} for the rate 

We proceed first by choosing specific emission directions for the particles. We select three directions of reference: 1) along the $\hat{y}-$axis; 2) along the $\hat{z}-$axis; 3) a direction characterized by polar and azimuth angles, respectively, $\theta=\ang{75}$ and $\phi=\ang{30}$ (hereafter we refer to this direction as ``\textit{generic}'' or ``\textit{G}''). For each direction, we have chosen the initial value of the magnetic field $B_0$ and, consequently, the maximum value of particles Lorentz factor $\gamma_0$.
Table~\ref{tab1} lists the values of $B_0$ and $\gamma_0$ for each emission direction and for the selected values of $\Upsilon$ that satisfy the condition in Eq.~\eqref{45}, and the one for a classical treatment of the problem (see section~\ref{Landaulevels}).

\begin{table}[!htb]
\centering
\caption{Maximum initial upper values for $B_0$ (in unit of the critical field $B_{cr}$) and $\gamma_0$, for the three initial emission directions of the particles, for the three selected values of $\Upsilon$, necessary in order to satisfy the condition given in Eq.~\eqref{45}.}
\label{tab1}
\begin{tabular}{lccc}
\hline
$\Upsilon$ & Direction & $B_0(B_{cr})$ & $\gamma_0$\\
\cline{1-4}
\multirow{3}*{$\frac{1}{2}$} & $y$ & $0.1$ & $3.66$ \\
\cline{2-4}
                 & $z$ & $0.1$ & $7.098$ \\
\cline{2-4}
                 & $Generic$ & $0.1$ & $6.48$ \\ 
\hline
\multirow{3}*{$\frac{1}{10}$} & $y$ & $0.1$ & $3.71$ \\
\cline{2-4}
                             & $z$ & $0.1$ & $22.66$ \\
\cline{2-4}
                             & $Generic$ & $0.1$ & $4.18$ \\   
\hline
\multirow{2}*{$\frac{1}{100}$} & $y$ & $0.1$ & $3.71$ \\
\cline{2-4}
                              & $Generic$ & $0.1$ & $3.81$ \\
\hline
\end{tabular}
\end{table}
For the values in Table~\ref{tab1}, we have integrated our system of equations varying the initial number of emitted particles, $N_{\pm,0}=1$, $10^3$, $10^6$, $10^{10}$, with $N_{\gamma,0}=0$; $N_{\gamma,0}=10^3$,  with $N_{\pm,0}=1$. Each numerical integration stops when $\gamma=1$, 
i.e. when the particle has lost all of its energy. 
We start the integration at $t_0=10^{-21}$~s and the previous condition is reached at $t_f \sim 3\times 10^{-17}$--$10^{-15}$~s, depending on the specific initial conditions. %

Figure~\ref{figBfield1} shows an appreciable decrease of $\vec{B}$ is obtained for high values of $N_{\pm,0}$ ($\geqslant 10^{10}$), with particles emitted along the $\hat{y}$ direction (as expected) and increasing $\Upsilon$. For particles emitted along the \textit{generic} direction, 
the screening increases for $\Upsilon=1/2\to 1/10$, while decreases for $\Upsilon= 1/10\to 1/100$. %

We obtain no exponential growth of the produced number of pairs, e.g. for $N_{\pm,0}=10^{10}$, only $10^2$--$10^3$ new pairs are created, and for $N_{\pm,0}=10^6$, only a few are created. This result tells us that the MPP process is not being efficient for all the cases in the time interval in which the particles lose their energy. When $N_{\pm,0}$ is high ($\sim 10^{10}$ or larger), the increase in the number of particles is mainly due to the larger number of photons rather than to a larger pair production rate.
%
\subsection{Magnetic field screening}
\label{subsecmagneticfieldresults}
\begin{figure}[ht!]
\centering
\includegraphics[width=\hsize,clip]{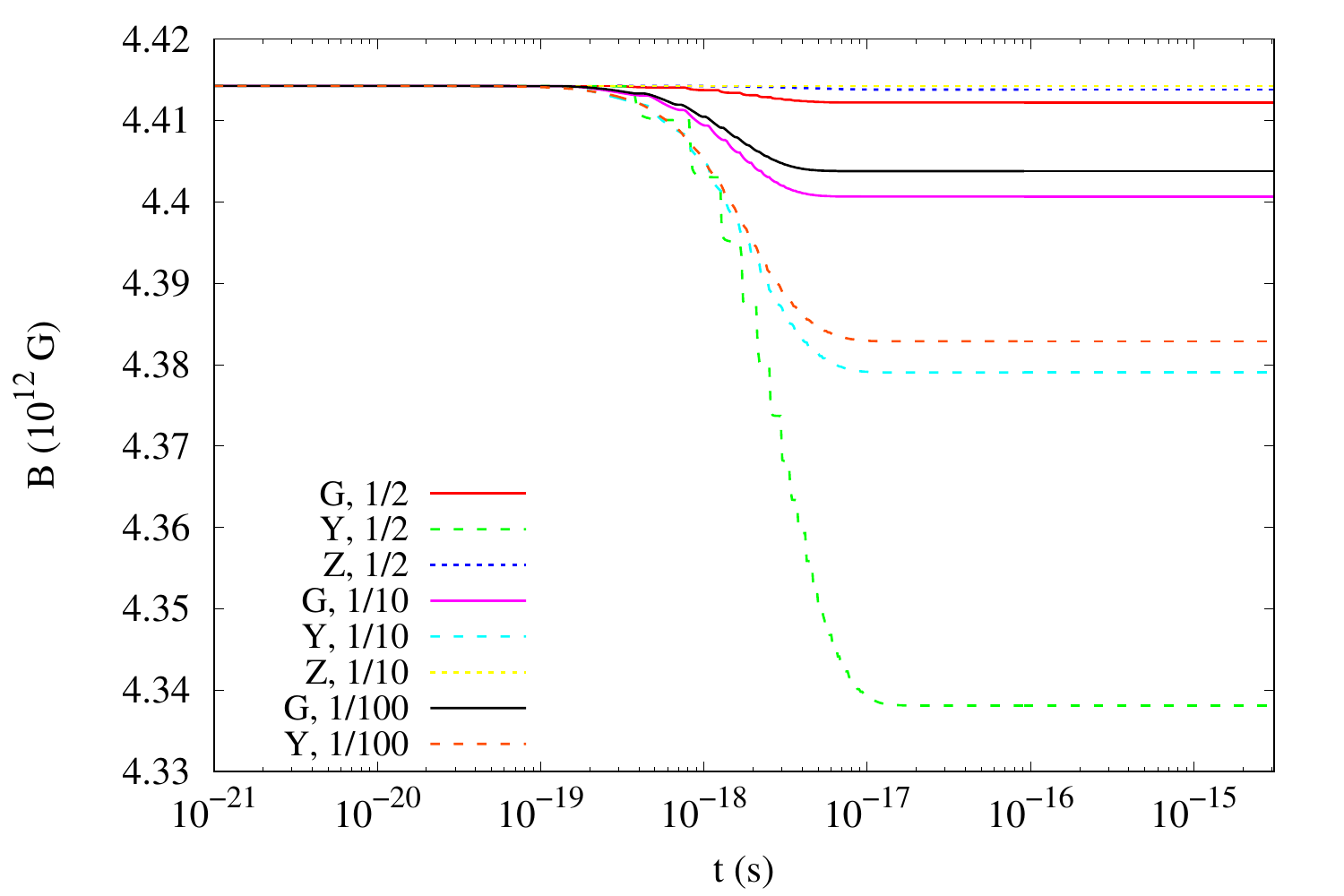}
\caption{The magnetic field decrease with time for the three values of $\Upsilon=1/2,~1/10,~1/100$ (and $B_0=0.1~B_{cr}$), operated by an initial number of particles $N_{\pm,0}=10^{10}$ emitted initially along the three directions ``\textit{generic}'', $\hat{y}$ and $\hat{z}$ (only for $\Upsilon=1/2$ and$~1/10$), is shown. For the case of emission along the $\hat{z}$ direction, the decrease cannot be appreciated because of the small magnitude of the decrease itself.}
\label{figBfield1}
\end{figure}

%
Figure~\ref{figBfield1} shows the screening of the magnetic field for $B_0=0.1~B_{cr}$, $N_{\pm,0}=10^{10}$ and different $\gamma_0$, operated by 
particles emitted initially:~1) for $\Upsilon=1/2$ and $1/10$, along the three directions \textit{generic}, $y$ and $z$; 
2) for $\Upsilon=1/100$, along the \textit{generic} and $\hat{y}$ directions\footnote{Since the integration time is not equal for all cases, we have extended a few solutions with their last constant value until the end time of the longer solution.}.

\begin{figure}[h!]
\centering
\includegraphics[width=\hsize,clip]{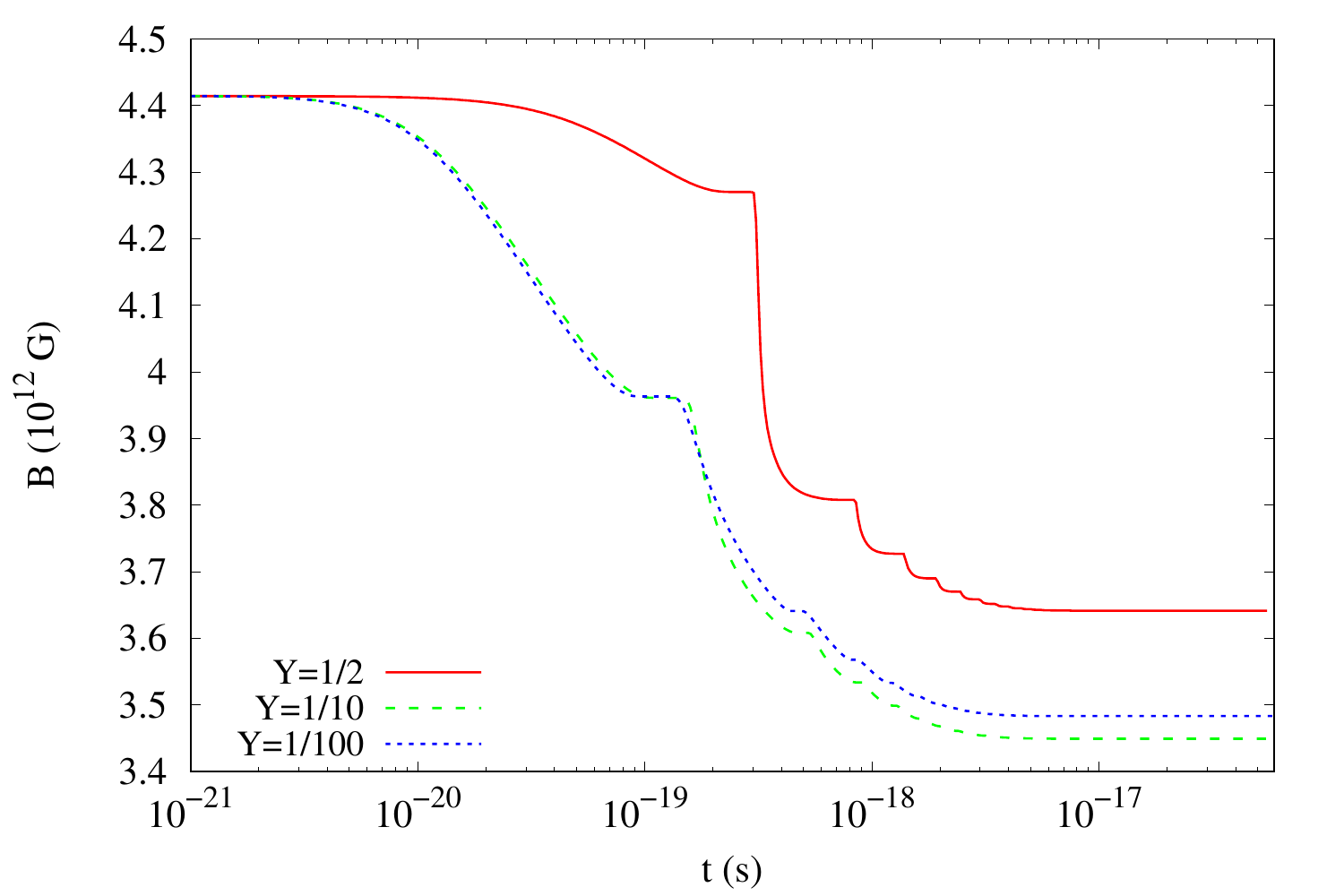}
\caption{The magnetic field decrease is shown, for $B_0=0.1~B_{cr}$ and $N_{\pm,0}=10^{15}$ emitted initially along the \textit{generic} direction, for $\Upsilon=1/2$, $1/10$, $1/100$, with Lorentz factor $\gamma_0=6.48$, $4.18$, $3.81$, respectively.}
\label{figBfield5}
\end{figure}

Figure~\ref{figBfield5} shows the screening of the magnetic field for $B_0=0.1~B_{cr}$, when $N_{\pm,0}=10^{15}$ particles are 
emitted along the \textit{generic} direction. The three curves correspond to the three values of $\Upsilon$. 

Figures~\ref{figBfield1} and~\ref{figBfield5} tell us that the larger the initial number of particles, the faster the magnetic field screening. It can be also seen that in all cases the screening process is stepwise (even if in some cases it is smoothed out) due to the dependence of Eq.~\eqref{22} on $\gamma$, $\beta_x$, and $\beta_y$, which have an oscillatory behavior owing to the continuous competition between gain and loss of energy.%
%

\subsection{Photons energy}
\label{subsecphoton}
\begin{figure}[ht!]
\centering
\includegraphics[width=\hsize,clip]{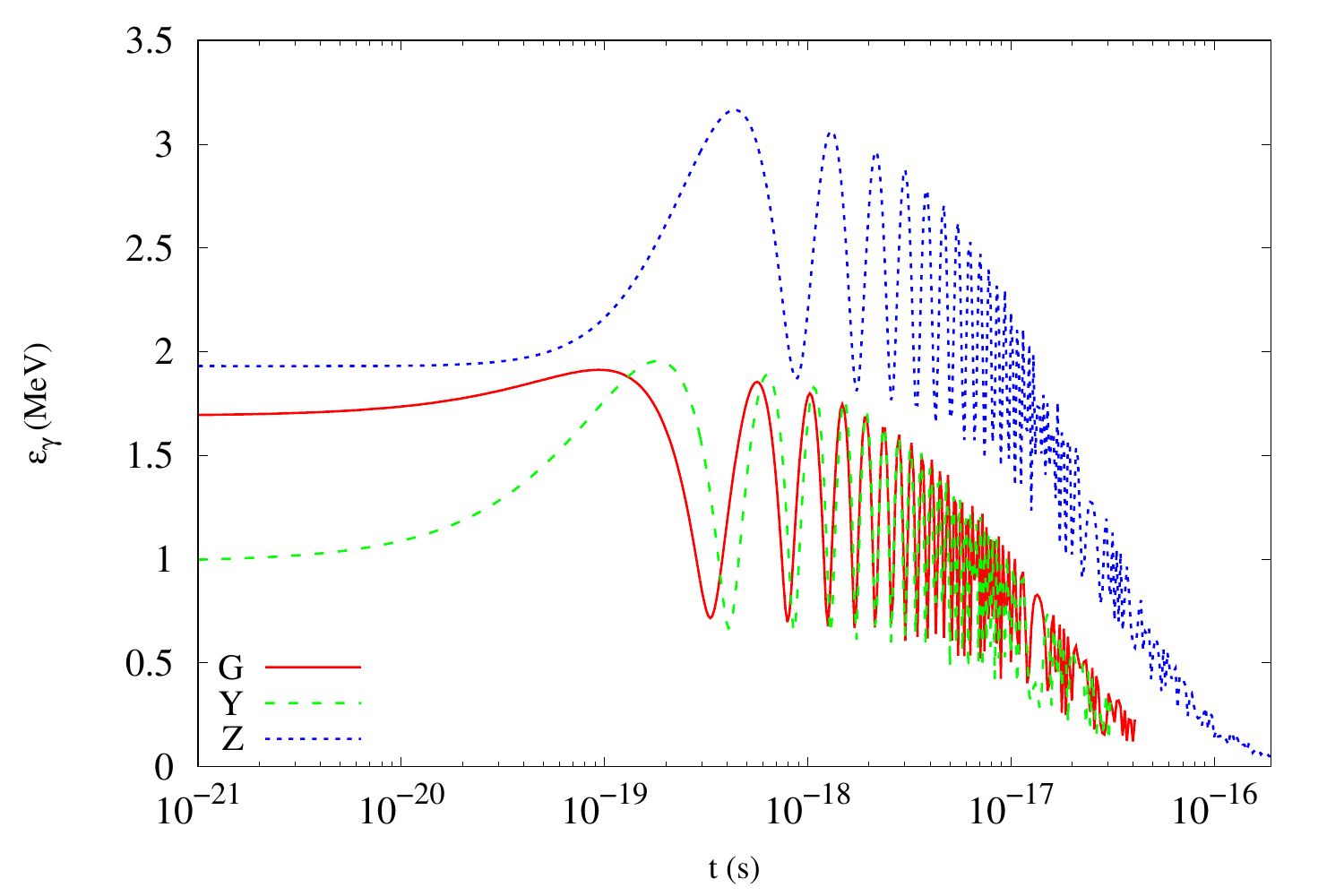}
\caption{Evolution of the photons energy for: $\Upsilon=1/2,~B_0=0.1~B_{cr}$ and $~N_{\pm,0}=10^{10}$ emitted along the three directions.}
\label{figPhoEne1}
\end{figure}

%
We here show the results for the photons energy and number. 
Figure~\ref{figPhoEne1} shows the photons energy $\varepsilon_{\gamma}(t)$ for $\Upsilon=1/2$, $B_0=0.1~B_{cr}$, $N_{\pm,0}=10^{10}$ and particles emitted in the three considered directions.
As before, the oscillatory behavior 
is due to the evolution of $\gamma$, $\beta_x$, $\beta_y$ 
that corresponds to a competition between acceleration of the particle (due to $\vec{E}$) and emission of radiation (due to $\vec{B}$). 

\begin{figure}[h!]
\centering
\includegraphics[width=\hsize,clip]{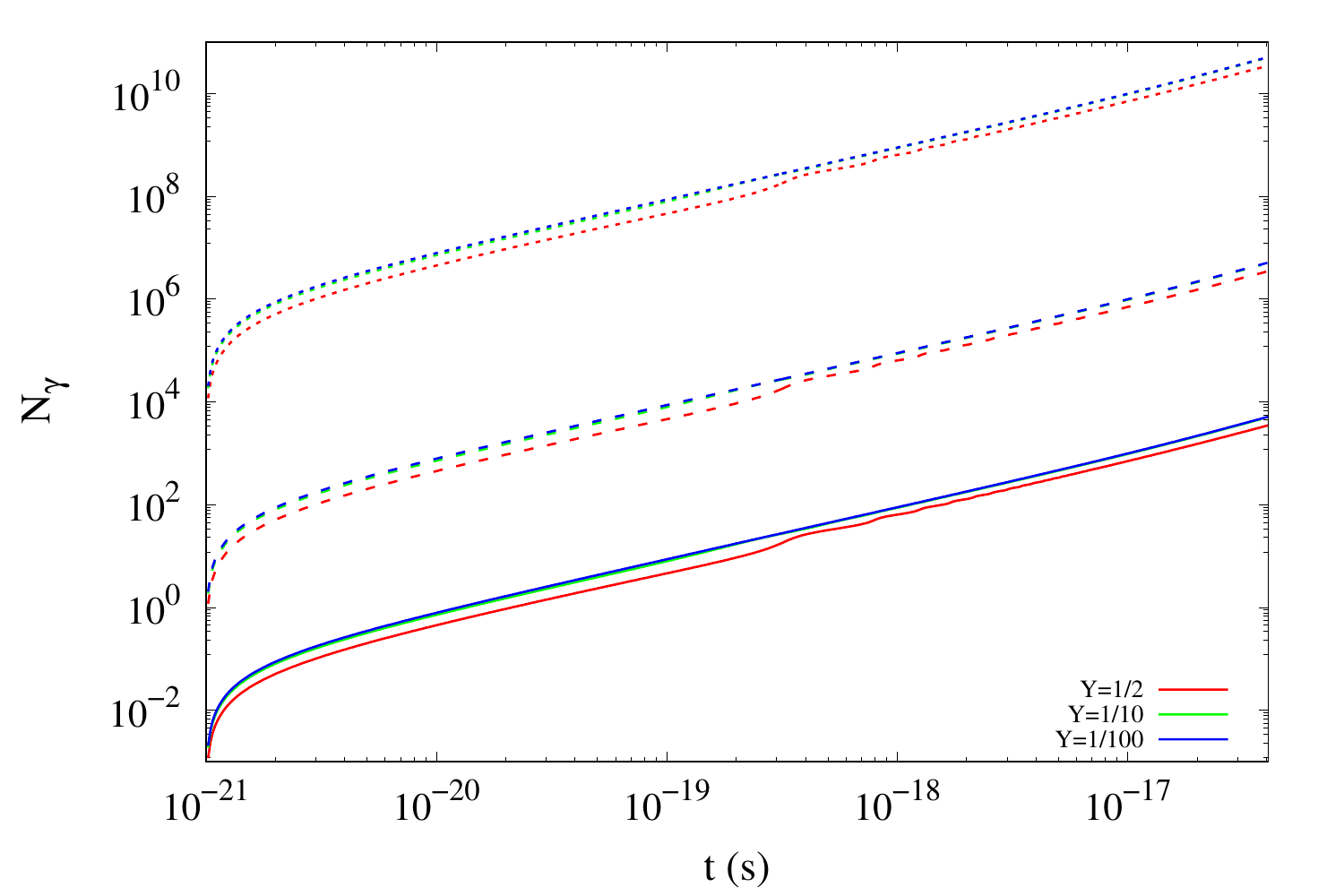}
\caption{Number of photons created by different values of the initial number of emitted particles $N_{\pm,0}=10^3$~(continuous lines),~$10^6$~(dashed lines),~$10^{10}$~(dotted lines). Here we consider particles emitted along the \textit{generic} direction, with $B_0=0.1~B_{cr}$, for $\Upsilon=1/2$ (red lines),$1/10$ (green lines),$1/100$ (blue lines). The curves for $\Upsilon=1/10$ and $1/100$ are almost overlapped.}
\label{figNphottot}
\end{figure}

Figure~\ref{figNphottot} shows the number of synchrotron photons created by 
$N_{\pm,0}=10^3$, $10^6$, $10^{10}$, emitted along the \textit{generic} direction, for the three values of $\Upsilon$. 
We notice that, for each $N_{\pm,0}$, a decrease of $\Upsilon$ leads to the creation of a larger number photons. Since to a decrease of $\Upsilon$ it corresponds a decrease of the electric to magnetic field ratio (see Eq.~\ref{42}), this implies that  
a larger number of synchrotron photons is produced hence a larger number of secondary pairs. Moreover, we notice that an exponential growth of $N_{\gamma}$ is present and their final value $N_{\gamma,f}$ is always one order of magnitude larger than $N_{\pm,f}$.
%
%
\subsection{Screening and circularization timescales}
\label{subsectimescales}
%
In order to use Eq.~\eqref{22} 
for the induced magnetic field, specific conditions on the processes time scales need to be satisfied. We define the circularization time as $t_c=2\pi R_c/(\beta\,c)$,
namely the time the particle spends to complete one ``\textit{orbit}'' around the magnetic field line\footnote{Here, we approximate the coil as perfectly circular due to the short timescale and since we are interested only in its order of magnitude.}. We define the screening timescale as $t_{screen}(t)=\left\lvert B(t)/\dot{B}(t)\right\rvert$.
For $t_c<t_{screen}$, the magnetic field can be considered stationary in the considered time interval, and we can use Eq.~\eqref{22}. 
For $t_{screen}\lesssim t_c$ instead, the assumptions of stationary field is no longer valid.

For all the studied ICs, we find that $t_c<t_{screen}$ or $\ll t_{screen}$. Instead for $N_{\pm,0}>10^{15}$, $t_{screen}$ becomes smaller than $t_c$ (even if for not all the integration time). Then, we exclude this IC from our study.

\subsection{Further conditions for the magnetic pair production}
\label{subsecfurtherMPPcond}
%
We turn to analyze the reason of the paucity that we find in the MPP process. The parameter $\chi$ has a twofold role: 1) it sets the energy of the emerging pairs; 2) it sets a threshold for the efficiency of the MPP process. If $0.01\lesssim\chi\leq 1$, the emerging pairs share equally the photon energy. Instead, if $\chi>1$ or $\gg 1$, one of the pairs tends to absorb almost all the energy of the photon, and the other takes the remaining energy (see \cite{DaughertyPair} for details). 
It has been shown that the pair production is not expected to occur with significant probability unless $\chi\gtrsim 0.1$ (see e.g.~\cite{DaughertyPair} and references therein). 
For all the ICs in Table~\ref{tab1}, 
$0.1< \chi<1$. Then, a production of pairs through the MPP process is expected and the emerging pairs share almost equally the parent photons energy.

A further rule-of-thumb condition for MPP was derived in \cite{sturrock1971model} (see also \cite{daugherty1975pair,harding1978curvature}), where it is shown that the pair production occurs whenever $    \varepsilon_{\gamma}\times B_{\perp}\gtrsim 10^{18.6}=3.98\times 10^{18}$,

with $\varepsilon_{\gamma}$ the photon energy and $B_{\perp}$ the perpendicular (to the photon propagation direction) component of the magnetic field. Inserting an electric field (perpendicular to $\vec{B}$), one has
\begin{equation}
\label{48}
\varepsilon_{\gamma}\times B \left[\left(\eta_x-\frac{E}{B}\right)^2+\eta_y^2\left(1-\frac{E^2}{B^2}\right)\right]^{1/2}\gtrsim 10^{18.6}.
\end{equation}
For all the analyzed cases, this condition 
is satisfied since it spans values between $10^{18}$ and $10^{23}$ (depending upon the ICs), even if not at all the integration times.

\section{Conditions for classical approach}
\label{Landaulevels}
%
We turn now to validate our semi-classical treatment of the screening problem. Quantum-mechanical effects are not important when the electron's cyclotron radius $R_L=cp/eB$ is larger than de Broglie wavelength $\lambda=\hslash/p$ (see~\cite{daugherty1976theory}), where $p=m\gamma\beta c$ is the electron's momentum. This corresponds to the following request for the magnetic field strength: $B\leq B_{\rm cr}~\beta^2~\gamma^2$. Moreover, in presence of an electric field $E$, the work exerted by the electric force over a de Broglie wavelength, $eE \lambda$, must be smaller than the electron's rest mass-energy, $mc^2$. This condition translates into $E<\gamma\beta E_{\rm cr}=\gamma \beta B_{\rm cr}$. For the parameters adopted in Table~\ref{tab1}, the above two conditions are well satisfied, so we do not expect the electrons in our system to experience quantum-mechanical effects, thereby validating the present semi-classical approach. In cases where the above conditions fail to be satisfied, e.g. in presence of overcritical fields, quantum-mechanical effects occur and the semi-classical approach for the dynamics and for the radiation production mechanisms are no longer valid. In those cases, the equation for the quantum synchrotron transitions rate suggested in~\cite{sokolov1968radiation} should be used. We here limit ourselves to physical situations in which the semi-classical treatment remain accurate (see Table~\ref{tab1}).

\begin{figure}[ht]
\centering
\includegraphics[width=\hsize,clip]{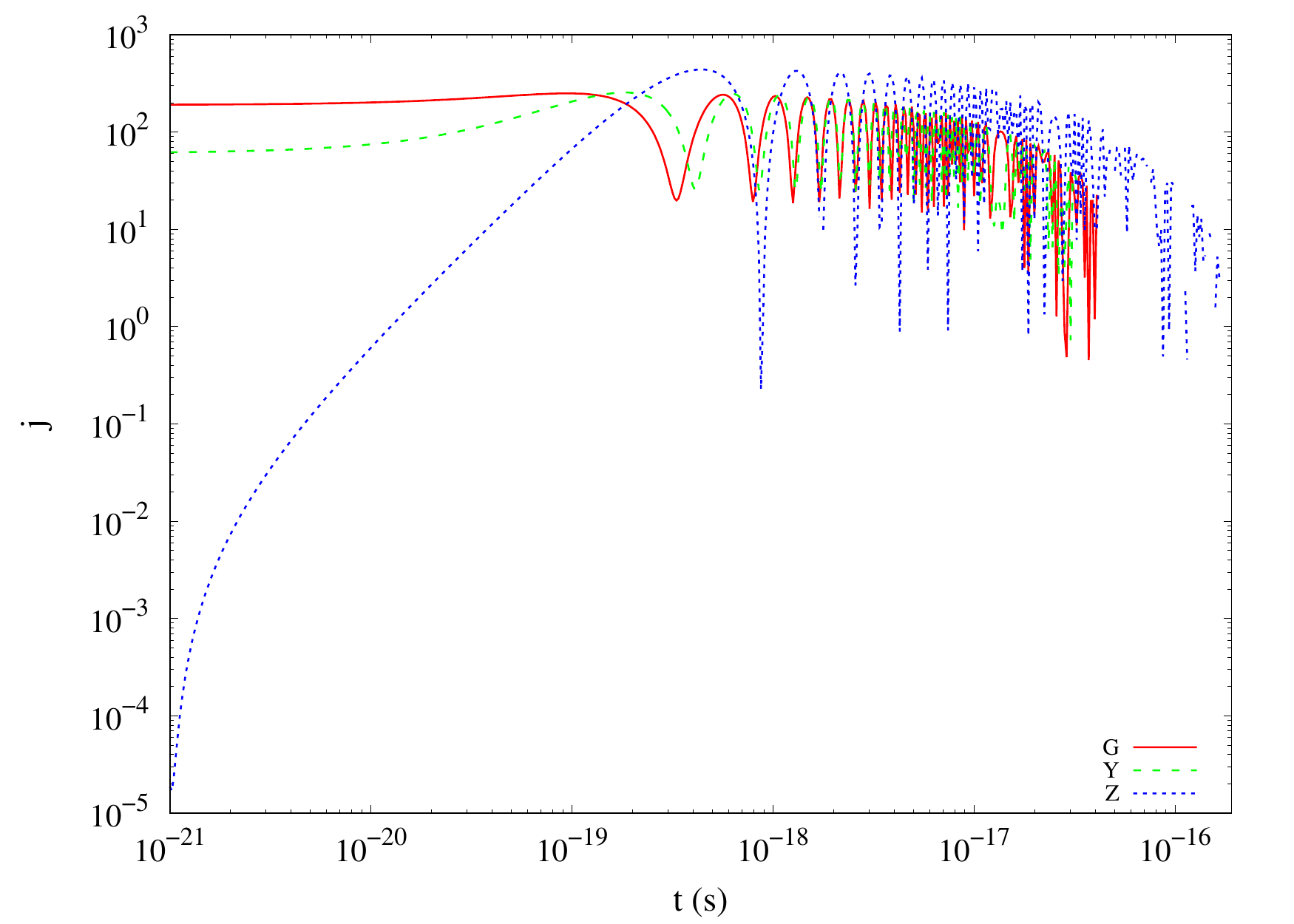}
\caption{Evolution of the Landau levels for the ICs (with~$B_0=0.1~B_{\rm cr},~N_{\pm}=10^{10}$ and particles emitted along the three directions $G$,~$y$ and~$z$) for $\Upsilon=1/2$. For~$\Upsilon=1/10$ and for the curves oriented along $G$ and $y$ directions,~with $\Upsilon=1/100$, the curves have almost the same behavior and values of the ones for $\Upsilon=1/2$.}
\label{figLL1}
\end{figure}

The above considerations can be also verified by looking at the particle's Landau levels. The energy of a particle immersed in strong background magnetic field is given by (see e.g. ~\cite{DaughertyPair})
\begin{equation}
\label{LL1}
E_j=\sqrt{\abs{\vec{p}_{\parallel}}^2 c^2 + m^2 c^4 +2 m^2 c^4 \frac{B}{B_{\rm cr}}~j},
\end{equation}
where $j$ is the number of occupied (Landau) energy levels and $p_{\parallel}$ its momentum component parallel to the magnetic field, $p_{\parallel}=p_z=\gamma m \beta_z c$. 
For given values of $\left(\gamma,~\beta_z,~B\right)$, from Eq.~\eqref{LL1} we can extract $j$:~$j= B_{\rm cr}/2 B\left[\gamma^2\left(1-\beta_z^2\right)-1\right]$.
The use of a classical treatment is allowed when the number of Landau levels is large, i.e. $\gg 1$. 
We show in Figure~\ref{figLL1} the value of $j$ as a function of time, for the ICs in Table~\ref{tab1}. High values of $j$ are reached for all the studied cases. We have found that $j\lesssim $ or $\ll 1$ for: 1) $\Upsilon=1/100$ and particles emitted along the $\hat{z}$ direction ($j$ oscillates between $10^{-3}-10^0$), and 2) $B_0>0.1 B_{\rm cr}$.
%
\section{Conclusions}
\label{sec8}
%
In this article, we have built a simplified model to study the MFS 
by $e^{\pm}$ pairs also in presence of a crossed electric field. 
Before to resume the results of our study, three important comments need to be considered about the model and the results obtained: 
\begin{enumerate}
\item 
We have constructed one-particle equations to describe the particles motion as a fluid. 
This assumption can be justified by the following considerations. Since we are considering strong fields, the particles are bound to follow almost the same trajectory.
Further, the flux of particles
can be treated as a fluid, since it obeys to the continuity equation. The particles flux along the lateral surface of the tube flux can be approximated to zero, while the ones through the upper and lower surfaces are equal since $e^-$ and $e^+$ move in opposite directions. 
\item
Because of Eq.~\eqref{42}, also the electric field is screened. This effect can be justified considering that the creation of new charged particles leads to the formation of a current which screens the electric field. The elaboration of a more detailed treatment of this phenomenon goes beyond the scope of the present article and is left for a future work.
\item 
The screening is mainly operated by the initial particles injected in the system. Under the studied conditions and time interval, the MPP is not sufficiently efficient.
In fact, 
the photons energy is of the order of a few MeV, so the $e^{\pm}$ pairs gain an energy just a bit higher than their rest-mass energy. As a consequence, they do not make many ``\textit{loops}'' around the $\vec{B}$ lines and emit photons with almost the same energy. This leads to a lower MPP rate. 
\end{enumerate}

We have shown that the screening increases (up to a few percent) if one increases the initial number of pairs, from $N_{\pm,0}=10^6$ to $10^{10}$--$10^{15}$.
It also depends on the initial direction of emission of the particles. The major effect occurs when the particles are emitted in the \textit{generic} and $\hat{y}$ directions, since the screening is produced by orthogonal component (respect to the $\hat{z}-$axis) of the particle velocity.

A further dependence is related to the parameter $\Upsilon$. A decrease of $\Upsilon$ enhances the efficiency of the screening since, because of Eq.~\eqref{42}, it leads to a decrease of the electric field strength. Consequently, the synchrotron process is more efficient and a higher number of photons is created. This also implies an increase of the MPP rate $\zeta(t)$. We can also notice the following features:
\begin{enumerate}
\item 
\textbf{Fixing $\Upsilon$:} the screening is larger if the particles are emitted initially along the $\hat{y}$-axis; it is lower if they are emitted along the \textit{generic} direction.
\item 
\textbf{Fixing the $\hat{y}$ direction:} the screening increases if we increase the value of $\Upsilon$.
\item 
\textbf{Fixing the \textit{generic} direction:} the screening increases if we decrease $\Upsilon$ (even if not linearly).
\end{enumerate}
The first feature is related to the particle orthogonal velocity, $\beta_{\perp}$, which is larger for particles emitted along the $\hat{y}$-axis, with respect to 
particles emitted along the \textit{generic} direction. 
The other two points are related to the dependence of the equation for the magnetic field, and in particular of the rate, on $\left(\vec{\beta},~\vec{\eta},~\Upsilon\right)$. 
Concerning the second point, we have verified that:~1) an increase of $\Upsilon$ leads to an increase of $\beta_{\perp}$;~2) in the time interval $10^{-21}\leq t \lesssim t^*=5\times 10^{-18}$~s, being $t^*$ the time when the magnetic field starts to drop down, the rate $\zeta(t)$ for $\Upsilon=1/10$ and $1/100$ is higher than the one for $\Upsilon=1/2$. For $t> t^*$, even if the rate for $\Upsilon=1/2$ is just little higher than for $\Upsilon=1/10$, $1/100$, it is
higher enough to explain a wider decrease of $\vec{B}$ for larger $\Upsilon$, for particles along the $\hat{y}$ direction.
This implies also a higher value for the respective $dN_{\pm}/dt$.
For the third point, analyzing Eq.~\eqref{22}, together with Eqs.~\eqref{31}, one can derive analytically that a decrease of $\Upsilon$ leads to a stronger MPP rate $\zeta(t)$. Moreover, a decrease of $\Upsilon$ implies a lower value for the particle Lorentz factor.  
In Figure~\ref{figNphottot}, we have also shown that a decrease of $\Upsilon$ leads to a stronger synchrotron emission, with the related increase of $N_{\gamma}$. Then, since $dN_{\pm}/d\tilde{t}= N_{\gamma}(\tilde{t})~\tilde{\zeta}(\tilde{t})$ and $d\tilde{B}_{tot}/d\tilde{t}\propto \gamma^{-2}\times dN_{\pm}/d\tilde{t}=\gamma^{-2}\times N_{\gamma}(\tilde{t})\times \tilde{\zeta}(\tilde{t})$, the discussions above imply that lower values of $\Upsilon$ leads to a stronger screening.

We conclude that the screening effect occurs under physical conditions reachable in extreme astrophysical systems, e.g. pulsars and gamma-ray bursts. For the present analyzed physical conditions, the decrease of the magnetic field from its original value can be of up to a few percent. This study has been the first one on this subject and in view of this, we have adopted some simplified assumptions that we have detailed and analyzed, and which have allowed us to get a clear insight on the main physical ingredients responsible for this effect. There is still room for improvements of the model, for instance, by considering different configuration of the electric and magnetic fields, overcritical fields strengths, among others. All the above considerations are essential to scrutinize the occurrence of the magnetic field screening process, and consequently for the interpretation of the astrophysical systems in which similar extreme physical conditions are at work.

\bibliography{biblio}
\end{document}